\documentclass[sigconf]{acmart}
\usepackage{booktabs}
\usepackage{multirow}
\usepackage{amsmath}

\newcommand{\myparatight}[1]{\smallskip\noindent{\bf {#1}:}~}

\usepackage[skip=2pt]{caption}

\graphicspath{{figures/}}

\newcommand{\mhj}{MHJ}
\newcommand{\wildchat}{WildChat}
\newcommand{\gradsafe}{GradSafe}

\copyrightyear{2026}
\acmYear{2026}
\setcopyright{cc}
\setcctype{by}
\acmConference[LAMPS '26]{3rd Workshop on Large AI Systems and Models with Privacy and Safety Analysis}{November 15--19, 2026}{The Hague, Netherlands}
\acmBooktitle{3rd Workshop on Large AI Systems and Models with Privacy and Safety Analysis (LAMPS '26), November 15--19, 2026, The Hague, Netherlands}
\acmDOI{10.1145/3846374.3846392}
\acmISBN{979-8-4007-3016-0/2026/11}

\begin{document}

\title[Does the Unsafe Gradient Survive a Conversation?]{Does the Unsafe Gradient Survive a Conversation? On the Fragility of Gradient-Based Jailbreak Detection in Multi-Turn Dialogue}

\author{Omar Sheta}
\email{omar.sheta@louisville.edu}
\affiliation{%
  \institution{University of Louisville}
  \city{Louisville}
  \state{Kentucky}
  \country{USA}}

\author{Rinku Deuja}
\email{rinku.deuja@louisville.edu}
\affiliation{%
  \institution{University of Louisville}
  \city{Louisville}
  \state{Kentucky}
  \country{USA}}

\author{Hadi Masoudi}
\email{mohammadhadi.masoudi@louisville.edu}
\affiliation{%
  \institution{University of Louisville}
  \city{Louisville}
  \state{Kentucky}
  \country{USA}}

\author{Minghong Fang}
\email{minghong.fang@louisville.edu}
\affiliation{%
  \institution{University of Louisville}
  \city{Louisville}
  \state{Kentucky}
  \country{USA}}

\begin{abstract}
Safety-aligned language models are commonly deployed as multi-turn assistants, allowing adversaries to distribute unsafe intent across several user turns rather than expressing it in a single prompt. Existing gradient-based jailbreak detectors, such as GradSafe, were developed for single-prompt inputs. They detect unsafe prompts by measuring the alignment between an input-induced gradient and a fixed unsafe reference direction; however, their effectiveness in multi-turn dialogue remains unclear. We conduct a controlled evaluation of gradient-based jailbreak detection in multi-turn settings. We extend GradSafe with a Context Window Scanner that applies the detector to fixed-size windows of user turns and uses the maximum window score as the conversation-level score. We evaluate different window sizes, attack families, benign conversation distributions, and target models. The results show a substantial difference between synthetic and realistic benign settings. When evaluated against synthetic benign conversations, the detector achieves an ROC-AUC of 0.98 against human-authored multi-turn jailbreaks. When evaluated on WildChat benign conversations, ROC-AUC decreases to 0.76, and a threshold calibrated on synthetic data incorrectly classifies more than 90\% of benign conversations as unsafe. Under realistic benign distributions, single-turn windows provide the highest separability, whereas longer windows and accumulated conversation contexts reduce performance. The detector is also sensitive to the attack-generation method and target model: successful Crescendo-generated attacks receive scores comparable to or lower than benign conversations, and evaluation on Qwen2.5-7B-Instruct yields near-random separability with a different optimal window size. These findings show that gradient-based signals can support multi-turn jailbreak detection, but reliable deployment requires calibration on realistic benign conversations, short-window scoring, length-aware thresholds, and evaluation across attack types and model architectures.
\end{abstract}




\maketitle

\section{Introduction}

Large language models are increasingly deployed as conversational systems rather than one-shot text generators. This shift changes the security problem. A malicious user no longer has to place the entire forbidden request in a single prompt; instead, they can distribute the objective across multiple apparently benign turns, establish context gradually, and ask a final question whose unsafe meaning depends on the preceding dialogue. Recent multi-turn attacks, including Crescendo-style escalation, iterative jailbreak generation, and human-authored multi-turn jailbreaks, exploit this structure by making intent emerge over time rather than appearing locally in one prompt~\cite{russinovich2025crescendo,chao2023pair,li2024mhj}.

Most jailbreak defenses, however, were developed for single-prompt settings and assume that the relevant signal is local. External classifiers such as Llama Guard label a prompt or a response against a safety taxonomy~\cite{inan2023llamaguard}, perturbation defenses such as SmoothLLM probe a prompt under random edits~\cite{robey2023smoothllm}, and internal-signal methods such as \gradsafe{}~\cite{xie2024gradsafe} read the gradient that a prompt induces when it is paired with a compliance token. Each of these was validated on isolated prompts, so none of them tells us how it behaves when the input is a conversation. This gap motivates a central question: when malicious intent is distributed across a dialogue, does the internal unsafe gradient signature remain detectable, weaken, or disappear?

We study this question by adapting \gradsafe{}~\cite{xie2024gradsafe} to multi-turn inputs through a Context Window Scanner (CWS), which slides a fixed-size window over the user turns, scores each window with \gradsafe{}, and reports the highest score, so that a single high-risk window can flag a conversation. This simple adaptation allows us to examine whether the original unsafe gradient signature remains separable as the amount of context, the realism of the benign data, and the attack family change. We contrast CWS with a full-history accumulated variant and a single-turn baseline, and evaluate all three methods on 537 human-authored multi-turn jailbreak (MHJ) conversations and 2{,}000 \wildchat{} benign conversations sampled to match the \mhj{} turn-length distribution.

Our results first show that detector performance depends strongly on the benign data. When benign conversations are synthetic, a three-turn window separates attacks almost perfectly and achieves an ROC-AUC of 0.98; when the same attacks are scored against realistic \wildchat{} conversations, the ROC-AUC falls to 0.76, and accumulated full-history scoring drops by a similar margin. This gap extends beyond ROC-AUC: the operating threshold chosen on synthetic data does not transfer, flagging more than 90\% of benign conversations when applied to real traffic and rendering the detector unusable in deployment. A benchmark built on clean synthetic benign prompts therefore overstates both accuracy and false-positive behavior.

When evaluated against realistic benign conversations, the study also overturns the intuition that more context helps. The single-turn baseline is the strongest setting overall, followed by a two-turn window, and separability declines steadily as the window grows or as the full history is accumulated. Length-conditioned and tactic-level analyses show the same trend: local windows remain strongest even on long conversations, and accumulated context is consistently weaker. We attribute this pattern to gradient dilution, because concatenating benign or unrelated turns adds noise to the gradient and pulls it away from the unsafe reference direction, so longer context masks rather than clarifies the signal.

The remaining findings concern generalization, and both are cautionary. When we replace human-authored attacks with automated Crescendo conversations, run on 200 HarmBench objectives and yielding 149 confirmed successes at a 74.5\% attack success rate, discriminative performance deteriorates substantially against the same \wildchat{} benign set: short windows approach random performance, and longer windows rank successful attacks below ordinary benign conversations. Because Crescendo was not optimized against the gradient score, this pattern reflects a structural mismatch rather than adaptivity: a successful context-building attack keeps its user turns close to benign gradients until the model generates the unsafe response. The same sensitivity appears across models. On Qwen2.5-7B-Instruct, separability is near random and the preferred window size reverses, with accumulated scoring becoming the strongest setting. This result shows that the behavior of a gradient detector depends on the model on which it is calibrated. Section~\ref{sec:crescendo-qual} analyzes the Crescendo mechanism in detail.

We summarize our contributions as follows.
\begin{list}{\labelitemi}{\leftmargin=1.2em \itemindent=-0.0em \itemsep=.2em}
\item We recast a single-turn internal-gradient detector, \gradsafe{}, as a multi-turn detector through a simple Context Window Scanner, and use it to test directly whether the unsafe gradient signature survives when intent is distributed across a dialogue.
\item We expose a large gap between synthetic and realistic benign calibration, in which synthetic benign data makes the detector appear nearly solved while realistic \wildchat{} traffic both lowers separability sharply and turns a synthetic-tuned threshold into severe over-blocking.
\item We show that shorter windows beat longer and accumulated context under realistic benign traffic, and we trace the effect to dilution of a sparse unsafe signal by benign dialogue.
\item We show that the detector does not generalize across attack families or architectures, since successful Crescendo attacks and a Qwen2.5-7B-Instruct target each collapse separability and the latter reverses the window-size preference.
\end{list}

\section{Background and Related Work}

\subsection{Single-Turn Jailbreak Attacks}

A jailbreak attack tries to make a safety-aligned model produce harmful or prohibited content. The earliest automated attacks worked at the level of a single prompt and showed that adversarial suffixes generalize across models. GCG~\cite{zou2023universal} searches for such a suffix with white-box gradients, maximizing the probability of an affirmative reply to a harmful request, and the suffixes it finds transfer to both open and proprietary systems. Tree of Attacks with Pruning~\cite{mehrotra2023tap} lowers the query cost by letting an attacker model expand and prune candidate prompts in a tree, so it reaches comparable success rates with far fewer queries than GCG. Studies of deployed systems show that the real threat surface is wider than optimized suffixes, because Shen et al.~\cite{shen2024doanything} catalog role-play, privilege-escalation, and prompt-injection patterns that ordinary users write without any gradient access. To make such attacks comparable, HarmBench~\cite{mazeika2024harmbench} and JailbreakBench~\cite{chao2024jailbreakbench} standardize the attacks, the target behaviors, and the refusal metrics, which is what allows reproducible evaluation across methods.

\subsection{Multi-Turn Jailbreak Attacks}

Multi-turn attacks move the threat from one crafted prompt to a coordinated dialogue. PAIR~\cite{chao2023pair} has an attacker model refine its jailbreak prompt through a feedback loop with a black-box target and succeeds in about twenty queries, treating each refinement as a step in an adversarial conversation. Crescendo~\cite{russinovich2025crescendo} instead escalates a sequence of innocuous requests, so that no single turn looks unsafe and the harmful content is only reached after enough shared context has been built. The MHJ dataset~\cite{li2024mhj} contains 537 human-written multi-turn jailbreak conversations organized by a tactic taxonomy spanning obfuscation, hidden-intention streamlining, injection, output formatting, request framing, echoing, and direct requests. It demonstrates that skilled humans exploit conversational coherence in ways that single-turn automation does not reproduce. Greshake et al.~\cite{greshake2023indirect} widen the surface further to indirect prompt injection, where instructions hidden in retrieved third-party content hijack the model inside an ongoing session without any explicit attacker turn. Taken together, these attacks argue that a detector must reason over dialogue history, since scoring a single prompt cannot capture intent that has been split across turns or planted in retrieved context.

\subsection{Jailbreak Defenses}

Existing defenses differ mainly in where they intervene. Input-level defenses act before the model runs: SmoothLLM~\cite{robey2023smoothllm} flags an attack when random mutations of the prompt change the model behavior, and perplexity filtering~\cite{jain2023baseline} uses the fact that GCG-style suffixes read as anomalously high perplexity, which gives a cheap signal without touching the weights. Classifier-level defenses add a second model: Llama Guard~\cite{inan2023llamaguard} treats safety as taxonomy classification over prompts and responses, and Llama Guard 3 ships with the Llama 3 family~\cite{grattafiori2024llama3} under a refreshed taxonomy. Internal-signal defenses instead read the target model itself. \gradsafe{}~\cite{xie2024gradsafe} observes that an unsafe prompt paired with a compliance token induces a characteristic gradient on safety-critical parameters, and scores a prompt by its cosine similarity to a fixed unsafe reference direction; SafeQuant~\cite{padakandla2025safequant} pursues the same idea with quantized gradients, and Gradient Cuff~\cite{hu2024gradientcuff} separates attacks by the gradient norm under a refusal-eliciting suffix without a stored reference. Gradient-Controlled Decoding~\cite{chiniya2026gcd} extends the single-anchor design to a pair of acceptance and refusal anchors and couples detection with a first-token mitigation step. 
Alignment-stage defenses modify model parameters during safety alignment or fine-tuning to improve robustness before deployment~\cite{gao2026neuronguard,huang2024antidote,chen2025vulnerability,yang2025alleviating}. They are complementary to our inference-time setting, which examines whether unsafe intent remains detectable when distributed across dialogue turns.
Our work stays within this internal-gradient family but asks the multi-turn question it has not addressed, namely whether the unsafe gradient remains readable when intent is distributed across a conversation. We do not use Gradient Cuff or Gradient-Controlled Decoding as baselines, because their refusal-loss and first-token mitigation protocols differ from our prompt-only, pre-generation setting, and we cite them only to place our study in context.
Remark that Jailbreak defense and forensic localization address different stages of the security process~\cite{zhang2025traceback,zhang2026Who,gao2026beware,gao2026patching}. A jailbreak defense identifies or prevents unsafe requests before the model produces harmful output. In contrast, forensic localization is performed after a safety failure to identify its cause.

\section{Threat Model}

\myparatight{Attacker}
The attacker reaches an aligned target model through a chat interface and controls a sequence of user turns, across which the malicious intent is deliberately distributed so that any one turn may read as benign, ambiguous, or incomplete. We grant the attacker awareness that a safety detector might be present, but not knowledge of the exact unsafe reference direction or the operating threshold.

\myparatight{Defender}
The defender aims to catch an unsafe conversation before the model emits harmful output. For \gradsafe{}-based scoring, the defender has the white-box access needed to compute gradients on safety-critical parameters. The detector is prompt-only, so it scores the user turns before generation and never sees the assistant responses. We choose this pre-generation setting on purpose, because once harmful content is produced it may already be seen, logged, cached, or copied, so intercepting the conversation before the first generated token is a stronger point of control. The point is sharper still in agentic and tool-use deployments, where a single response can launch irreversible actions such as external API calls, code execution, file writes, or downstream tool invocations, before any response-side moderator has a chance to act. The cost of this choice is that the detector cannot rely on evidence that only the response would reveal.

\myparatight{Scope}
We study the separability and threshold behavior of the detector, not a complete blocking system, and we do not claim robustness against an attacker that optimizes directly against the gradient score. In principle, an attacker who knew the unsafe reference direction could steer a dialogue to lower its cosine similarity while keeping the harmful goal, but in a multi-turn setting that optimization would still have to preserve coherent context and elicit unsafe behavior, and it lies outside our scope. For fairness, the Llama Guard baseline is run under the same prompt-only, user-turn-only protocol, which is narrower than its intended use as a prompt and response safeguard.

\section{Our Method}

We build our detector by reusing the scoring rule of \gradsafe{}~\cite{xie2024gradsafe} unchanged and adding only the machinery needed to apply it to a conversation. The separation is deliberate, because it keeps the single-turn scoring rule fixed so that any change in behavior can be traced to how context is assembled rather than to a new detector design. We first restate the scoring rule and then describe the windowing procedure that turns it into a multi-turn detector.

\subsection{GradSafe Scoring}

\gradsafe{} builds on the observation that a harmful request leaves a consistent trace in the gradient of a safety-aligned model. When an input is paired with a short compliance token such as ``Sure,'' an unsafe input drives a small set of safety-critical parameters in a direction that is stable across different unsafe prompts and separable from the direction that benign inputs induce. \gradsafe{} turns this into a detector by fixing that direction once, as an unsafe reference signature $\mathbf{g}_{\mathrm{unsafe}}$ averaged over a set of unsafe seed prompts, and by restricting attention to the safety-critical parameters $\theta_{\mathcal{S}}$ on which the separation is sharpest. Writing $x$ for the detector input and $t$ for the compliance token, the score is the cosine similarity between the induced gradient and this reference signature,
\begin{equation}
S(x) =
\cos\left(
\nabla_{\theta_{\mathcal{S}}}\mathcal{L}(x \oplus t, t),
\mathbf{g}_{\mathrm{unsafe}}
\right),
\end{equation}
so a larger score means closer alignment with the unsafe reference. The score is a single scalar and needs one backward pass rather than any text generation, which is what allows it to act as a pre-generation filter. In the implementation the loss is taken over the full compliance suffix that follows the prompt, which is the separator marker followed by ``Sure,'' and the reference signature and every evaluated score use this same suffix so that the comparison stays consistent.

\subsection{Context Window Scanner}

A conversation is a sequence of turns, while the scoring rule above consumes a single input, so a multi-turn detector must decide which text to score. The two obvious choices are both unsatisfying. Scoring only the most recent user turn discards the context that a multi-turn attack is designed to accumulate, whereas scoring the whole history at once blends the attack turn with every benign turn around it. The Context Window Scanner (CWS) spans these two choices with one parameter, the window size $W$, and keeps everything else minimal.

Because the defender is prompt-only, CWS first drops the assistant turns and keeps the user turns,
\begin{equation}
H = \langle h_1, h_2, \ldots, h_m \rangle .
\end{equation}
It then slides a right-anchored window of size $W$ that ends at each user turn,
\begin{equation}
\begin{aligned}
x_{t,W}
&= \mathrm{join}_{\delta}(h_{\max(1,t-W+1)},\ldots,h_t),
\end{aligned}
\end{equation}
for $t=1,\ldots,m$, so the window is causal and could be evaluated online as each turn arrives, and the first $W-1$ windows are shorter simply because no earlier turns exist. Adjacent turns inside a window are joined by a fixed delimiter $\delta$, realized as a blank-line-separated horizontal rule, which keeps the turn boundaries explicit in the scored text. \gradsafe{} scores each window, and the conversation takes the largest window score,
\begin{equation}
S_{\mathrm{CWS},W}(H) = \max_{1\le t\le m} S(x_{t,W}).
\end{equation}
Max pooling is a design choice rather than a convenience, because it treats the windows as a logical OR in which one high-risk window is enough to flag the conversation, which matches a deployment where a single successful jailbreak turn already counts as a failure.

The window size directly controls how much context enters each score. At $W=1$ the scanner reduces to scoring each user turn in isolation, and as $W$ grows it incorporates progressively more history, until a single window over all turns recovers full-history accumulation. Thus, the single-turn and accumulated baselines represent the two endpoints of the context-length spectrum. Scoring a conversation of $m$ user turns costs $m$ backward passes, one per window position and independent of $W$, so the procedure remains inexpensive; the main risk of a larger window is mixing benign turns into the scored text. We intentionally keep CWS simple. It is not proposed as a complete defense but as a faithful extension of a single-turn detector to the multi-turn setting, and its simplicity allows the experiments to attribute each effect to context length, benign-data realism, or attack family rather than to a more elaborate detector.

\section{Experimental Setup}

\myparatight{Attack data}
The attack set is the 537 \mhj{} conversations~\cite{li2024mhj} taken from the bundled HarmBench behavior file, where each row is a multi-turn user sequence tied to a harmful behavior or tactic.

We add a second, automated attack family by running Crescendo on 200 HarmBench objectives. The pipeline uses PyRIT Crescendo with a local 70B attacker model against the Llama 3.1 8B target endpoint, capped at 10 turns, 10 backtracks, and a 5-second gap between objectives. Rather than keyword matching, success is judged by a calibrated scorer that rates each target response on a scalar scale and converts the rating to a Boolean label at a threshold of 0.9. 
All 200 objectives completed, of which 149 were confirmed successful, giving a 74.5\% attack success rate, and the successful conversations average 3.13 user turns, with median 3, minimum 1, and maximum 7. The other 51 rows failed at runtime because the attacker prompt exceeded the 8{,}192-token endpoint limit, and we do not count them as attack failures; since no completed target transcript survives for them, the all-attempt set stores only the original HarmBench objective as one user prompt.

For all main claims, we use the 149 confirmed successes as the successful-attack subset and report the 200-row all-attempt set only for provenance. This distinction is important because each of the 51 fallback rows reduces to a single-turn, explicit harmful request, which produces a strong localized unsafe gradient and bypasses the context dilution under study. Including these rows increases the single-turn ROC-AUC from a near-random $0.5018$ on confirmed successes to $0.5856$ on all attempts. We therefore exclude them and restrict our conclusions to genuine multi-turn context-building attacks.

\myparatight{Benign data}
Because false positives decide whether a prompt-only detector is deployable, we calibrate on realistic traffic rather than clean synthetic prompts. The primary benign benchmark is a stratified sample of 2{,}000 \wildchat{}~\cite{zhao2024wildchat} conversations, a corpus of real interaction logs whose in-the-wild traffic already mixes genuine jailbreaks with benign but safety-relevant discussion~\cite{jiang2024wildteaming}. The sampler first drops one-turn conversations and then matches the \mhj{} turn-length distribution over the remainder, yielding 740 short conversations of two or three user turns, 680 medium conversations, and 580 long conversations.

\myparatight{Synthetic benign data}
We also keep the early synthetic benign benchmark from the initial evaluation, 183 benign multi-turn conversations generated with Gemini 2.5 Pro over ordinary assistant topics such as cooking, travel, hobbies, general knowledge, and short writing. These are mostly two turns long, uniformly polished and on-topic, and free of the emotionally intense, adversarial-looking, or policy-sensitive content that fills real traffic. As a stand-in for the clean synthetic prompts that earlier single-turn evaluations rely on, it serves to quantify the evaluation gap rather than to support any deployment claim. 
Two properties make these conversations easy to separate from attacks: their clean content and their short, unmatched length distribution. Consequently, this benchmark overstates performance along both realism and length axes, so we treat the length-matched WildChat benchmark as the realistic reference throughout.

\myparatight{Models}
The main experiments use Llama 3.1 8B Instruct~\cite{grattafiori2024llama3}, and we add Qwen2.5-7B-Instruct~\cite{qwen2025report} as a cross-architecture target. The Qwen2.5 reference signature is built with the same procedure as the Llama 3.1 signature, using a separately optimized safety-critical parameter mask over the same unsafe seed prompts, and the cross-model results appear in Section~\ref{sec:cross-model}.

\myparatight{Baselines}
We compare CWS against the two extremes it interpolates and against an external classifier. The single-turn max-over-turn baseline scores each user turn on its own and keeps the highest score, which is the $W=1$ point and the reference for whether added context helps at all. The accumulated baseline concatenates every user turn into one input and scores it once, which tests whether the full history preserves or dilutes the unsafe signature. Finally, we run Llama Guard 3 (\texttt{meta-llama/Llama-Guard-3-8B}) with its official chat template under the same prompt-only, user-turn-only protocol, scoring and max-pooling each user-turn window for $W=3$ and scoring the concatenated user turns once for the accumulated mode, and we never include assistant responses or harmful outputs.

\myparatight{Metrics}
We report ROC-AUC, average precision (AP), the best-F1 threshold, best F1, and the true positive rate at a 10\% false positive rate. AP follows the project convention of averaging precision at each positive rank after scores are sorted in descending order, which matches the artifact script and can differ slightly from library implementations under ties. To test portability, we also transfer the synthetic-benchmark threshold to real \wildchat{} traffic and measure the resulting false-positive rate. Confidence intervals are nonparametric 95\% bootstrap intervals with 2{,}000 resamples; attack and benign conversations are resampled separately, and the window-ablation deltas use paired resampling over the shared rows. Rows whose scoring failed are excluded from the metrics and reported as errors where present.

\myparatight{Hardware and determinism}
Gradient scoring runs in bfloat16 on NVIDIA H100 GPUs with automatic device mapping. The stratified \wildchat{} sampler uses seed 13, the benign-subspace pilot uses seed 17, the bootstrap intervals use 2{,}000 resamples with seed 20260514, and the paired window-difference intervals use seed 20260515. The artifact script records the provenance of every table and figure, and code together with redacted data artifacts will be released upon acceptance, subject to the review policy and to responsible handling of harmful transcripts.

\section{Experimental Results}

\subsection{Synthetic versus Real Benign Calibration}

The clearest result of the study is that the benign data alone decides how strong the detector appears, and Table~\ref{tab:synthetic-real} makes the gap concrete. Against synthetic benign conversations both the three-turn window and accumulated scoring clear 0.97 ROC-AUC, but against the real stratified \wildchat{} set the same two settings fall to roughly 0.76 and 0.74, with average precision dropping in step.

\begin{table}[t]
\centering
\small
\caption{Synthetic benign data overstates detector performance. Threshold transfer applies the synthetic best-F1 threshold to the matching \wildchat{} setting.}
\label{tab:synthetic-real}
\resizebox{\columnwidth}{!}{
\begin{tabular}{llrrlrr}
\toprule
Benign set & Mode & AUC & Transfer FPR & 95\% CI & AP & F1 \\
\midrule
Synthetic & $W=3$ & 0.9836 & 0.0874 & [0.9750, 0.9909] & 0.9949 & 0.9740 \\
\wildchat{} & $W=3$ & 0.7614 & 0.9085 & [0.7379, 0.7843] & 0.4278 & 0.5260 \\
Synthetic & Accum. & 0.9777 & 0.1530 & [0.9675, 0.9864] & 0.9929 & 0.9576 \\
\wildchat{} & Accum. & 0.7376 & 0.9065 & [0.7130, 0.7614] & 0.4043 & 0.5039 \\
\bottomrule
\end{tabular}}
	\vspace{-.10in}
\end{table}

The threshold transfer numbers matter even more than the AUC drop. A three-turn threshold that flags only 8.74\% of synthetic benign conversations flags 90.85\% of real ones, and the accumulated threshold behaves the same way, so the synthetic benchmark does not merely inflate a score, it hands the practitioner an operating point that would reject almost all legitimate traffic.

\subsection{Main Detector Comparison}

On the realistic benchmark the ordering of the methods is clear, and Table~\ref{tab:main-results} reports it. The strongest \gradsafe{} setting is the single-turn window at 0.8101 ROC-AUC, the two-turn window comes next and still beats three turns, three turns remain usable but weaker, and accumulated full-history scoring is weakest of all, so shorter windows win across the board.

\begin{table}[t]
\centering
\small
\caption{Main results on \mhj{} attacks versus stratified \wildchat{} benign conversations.}
\label{tab:main-results}
\resizebox{\columnwidth}{!}{
\begin{tabular}{lrlrrr}
\toprule
Method & ROC-AUC & 95\% CI & AP & Best F1 & TPR@10\%FPR \\
\midrule
\gradsafe{} $W=1$ & 0.8101 & [0.7895, 0.8311] & 0.5135 & 0.5749 & 0.5084 \\
\gradsafe{} $W=2$ & 0.7873 & [0.7654, 0.8092] & 0.4647 & 0.5536 & 0.4227 \\
\gradsafe{} $W=3$ & 0.7614 & [0.7379, 0.7843] & 0.4278 & 0.5260 & 0.4004 \\
\gradsafe{} Accum. & 0.7376 & [0.7130, 0.7614] & 0.4043 & 0.5039 & 0.3203 \\
Llama Guard $W=3$ & 0.4528 & [0.4266, 0.4796] & 0.2118 & 0.3494 & 0.0838 \\
Llama Guard Accum. & 0.4409 & [0.4132, 0.4692] & 0.2036 & 0.3494 & 0.0615 \\
\bottomrule
\end{tabular}}
\end{table}

\begin{figure}[t]
  \centering
  \includegraphics[width=\columnwidth]{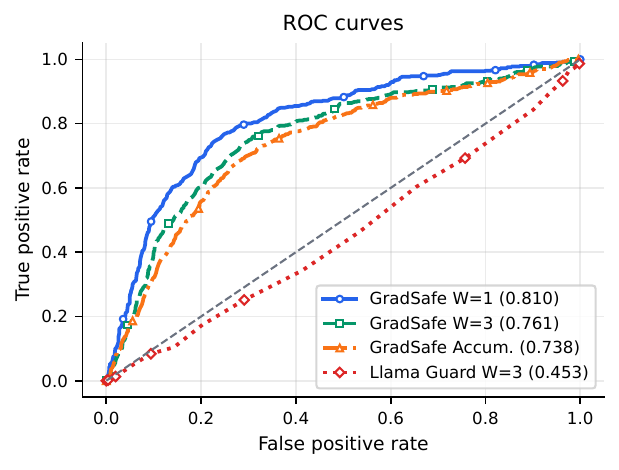}
  \Description{ROC curves for GradSafe $W=1$, GradSafe $W=3$, GradSafe accumulated, Llama Guard $W=3$, and Llama Guard accumulated on MHJ versus WildChat.}
  \caption{ROC curves for key methods on the real \wildchat{} benchmark.}
  \label{fig:roc}
\end{figure}

\begin{figure}[t]
  \centering
  \includegraphics[width=\columnwidth]{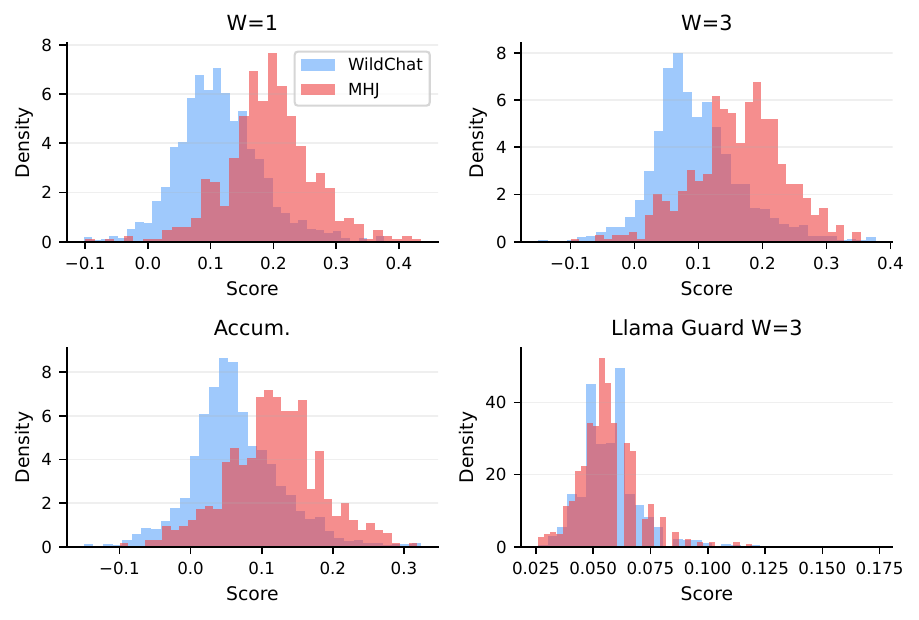}
  \Description{Density plots of attack and benign detector scores for representative GradSafe settings and Llama Guard, showing substantial overlap on real benign conversations.}
  \caption{Score distributions for representative settings. Real benign scores overlap heavily with attack scores.}
  \label{fig:score-distributions}
\end{figure}

Llama Guard 3 sits below 0.5 AUC in both modes, which we do not read as a general weakness of Llama Guard but as a mismatch of task. Taxonomy-based prompt classification is not built to catch intent that is still hidden before generation, so its failure here confirms that prompt-only gradient scoring occupies a different point in the pipeline from response-side moderation.

\myparatight{WildChat contamination}
A manual check of the stratified \wildchat{} sample turned up conversations that carry jailbreak templates, role-play safety tests, or other adversarial-looking text. We keep these rows in the benchmark rather than remove them, because they are a genuine part of real traffic and a genuine source of false-positive pressure for a prompt-only detector, so the unfiltered distribution, not a sanitized subset, is what drives our headline numbers.

\myparatight{Window size ablation}
Sweeping the window size confirms the trend seen above, and Table~\ref{tab:window-ablation} lays it out: AUC falls monotonically from one turn to seven, and accumulated scoring lands near the seven-turn window and well behind the short ones.

\begin{table}[t]
\centering
\small
\caption{Window ablation on \mhj{} versus \wildchat{}. Short local windows beat long windows and accumulated context. Deltas are paired bootstrap differences against $W=3$.}
\label{tab:window-ablation}
\resizebox{\columnwidth}{!}{
\begin{tabular}{lrlrlrr}
\toprule
Method & AUC & 95\% CI & $\Delta$ vs $W=3$ & 95\% CI & AP & F1 \\
\midrule
$W=1$ & 0.8101 & [0.7895, 0.8311] & 0.0487 & [0.0329, 0.0643] & 0.5135 & 0.5749 \\
$W=2$ & 0.7873 & [0.7654, 0.8092] & 0.0258 & [0.0155, 0.0365] & 0.4647 & 0.5536 \\
$W=3$ & 0.7614 & [0.7379, 0.7843] & 0.0000 & [0.0000, 0.0000] & 0.4278 & 0.5260 \\
$W=4$ & 0.7569 & [0.7325, 0.7799] & -0.0045 & [-0.0103, 0.0011] & 0.4150 & 0.5207 \\
$W=5$ & 0.7503 & [0.7261, 0.7734] & -0.0111 & [-0.0181, -0.0039] & 0.4057 & 0.5150 \\
$W=7$ & 0.7362 & [0.7119, 0.7594] & -0.0252 & [-0.0347, -0.0159] & 0.3849 & 0.4993 \\
Accum. & 0.7376 & [0.7130, 0.7614] & -0.0238 & [-0.0378, -0.0095] & 0.4043 & 0.5039 \\
\bottomrule
\end{tabular}}
\end{table}

The direction is the opposite of what one would expect if multi-turn detection needed more context. Here a short window isolates the local unsafe cue better than a long one, and the effect is statistically clean, since the paired interval for one turn against three turns is strictly positive at $\Delta$AUC $=0.0487$ with 95\% CI [0.0329, 0.0643]. The most consistent explanation is dilution, because each extra benign or unrelated turn adds gradient that is not aligned with the unsafe reference and therefore washes the signal out.

\begin{figure}[t]
  \centering
  \includegraphics[width=\columnwidth]{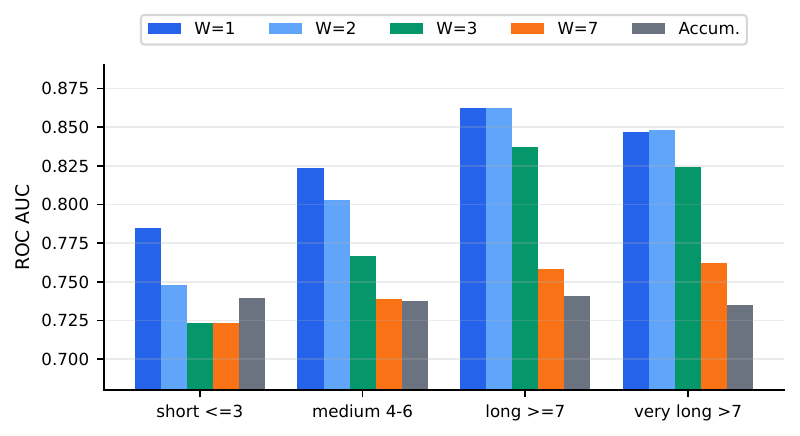}
  \Description{Grouped bar chart comparing ROC AUC by conversation-length bucket for $W=1$, $W=2$, $W=3$, $W=7$, and accumulated scoring.}
  \caption{Length-conditioned AUC. Short windows are strongest even on long conversations, while accumulated context stays weaker.}
  \label{fig:length}
\end{figure}

\myparatight{Length-conditioned behavior}
Because dilution should depend on how much context is added, we also break the results down by conversation length in Figure~\ref{fig:length}. Short windows dominate globally and, tellingly, stay dominant on the longest conversations: for conversations with seven or more user turns, the two-turn and one-turn windows reach 0.8625 and 0.8622 AUC while three turns reach 0.8370, and accumulated scoring trails at 0.7411.
This resolves the role of context in a practical way. A long conversation does not call for scoring its whole history; if anything, the extra length only creates more room for dilution, so a deployed detector should pair short-window scoring with length-aware threshold calibration instead of assuming that accumulating context is the safe default.

\begin{table}[t]
\centering
\small
\caption{Tactic-level AUC. Each tactic subset is scored against the full \wildchat{} benign distribution.\protect\footnotemark}
\label{tab:tactic}
\resizebox{\columnwidth}{!}{
\begin{tabular}{lrrrr}
\toprule
Tactic & $n$ & $W=1$ & $W=3$ & $W=7$ \\
\midrule
Direct Request & 146 & 0.7616 & 0.7277 & 0.7297 \\
Hidden Intention Streamline & 109 & 0.8006 & 0.7538 & 0.7014 \\
Injection & 32 & 0.9044 & 0.8727 & 0.8648 \\
Obfuscation & 156 & 0.7905 & 0.7235 & 0.6948 \\
Output Format & 23 & 0.8895 & 0.8641 & 0.8273 \\
Request Framing & 68 & 0.8980 & 0.8388 & 0.8027 \\
\bottomrule
\end{tabular}}
\end{table}
\footnotetext{The \mhj{} Echoing tactic has only three attack examples, so we omit it from the table and do not treat it as a trend.}

\myparatight{Tactic-level behavior}
The advantage of short windows also holds tactic by tactic, as Table~\ref{tab:tactic} shows. The one-turn window is strongest on every nontrivial tactic group, including direct requests, hidden-intention streamlining, injection, obfuscation, output-format attacks, and request framing.
This result does not imply that a single turn contains the entire malicious objective. Instead, it suggests that the signature responds most strongly to the local turns closest to the unsafe goal, while the surrounding context tends to dilute that signal rather than sharpen it.

\begin{figure}[t]
  \centering
  \includegraphics[width=\columnwidth]{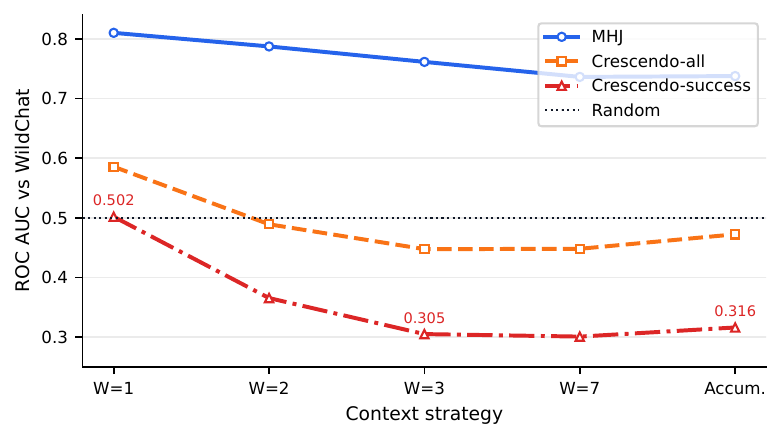}
  \Description{Line chart comparing GradSafe ROC AUC across MHJ, all Crescendo attempts, and successful Crescendo attacks. MHJ is highest, while successful Crescendo is near or below random.}
  \caption{Attack-family AUC. \gradsafe{} separates \mhj{} from \wildchat{}, but successful Crescendo conversations rank near or below benign traffic.}
  \label{fig:attack-family}
\end{figure}

\myparatight{Sensitivity to crescendo attacks}
The preceding results are measured on human-authored \mhj{} attacks and do not generalize to automated Crescendo attacks, as Table~\ref{tab:attack-family} and Figure~\ref{fig:attack-family} show against the same \wildchat{} benign set. Whereas \mhj{} achieves AUCs of 0.8101 and 0.7614 at one and three turns, respectively, successful Crescendo reduces the one-turn AUC to 0.5018 and pushes the three-turn, seven-turn, and accumulated settings well below 0.5.

\begin{table}[t]
\centering
\small
\caption{Attack-family sensitivity of \gradsafe{} against the stratified \wildchat{} benign set. Crescendo-success is the 149 confirmed successes; Crescendo-all adds 51 single-prompt objective-fallback rows.}
\label{tab:attack-family}
\resizebox{\columnwidth}{!}{
\begin{tabular}{lrrrrrr}
\toprule
Attack family & $n$ & $W=1$ & $W=2$ & $W=3$ & $W=7$ & Accum. \\
\midrule
\mhj{} & 537 & 0.8101 & 0.7873 & 0.7614 & 0.7362 & 0.7376 \\
Crescendo-all & 200 & 0.5856 & 0.4891 & 0.4474 & 0.4480 & 0.4723 \\
Crescendo-success & 149 & 0.5018 & 0.3654 & 0.3050 & 0.3008 & 0.3161 \\
\bottomrule
\end{tabular}}
\end{table}

This reduction is not statistical variation around random performance. The three-turn interval for successful Crescendo lies entirely below random ranking at a 95\% CI of [0.2608, 0.3503], and even the one-turn window is statistically indistinguishable from random at a 95\% CI of [0.4482, 0.5530].

A below-random AUC should not be interpreted as a useful inverted detector. Under this reference signature and prompt-only protocol, the user turns of a successful Crescendo attack receive lower scores than ordinary \wildchat{} traffic. Thus, the result is more serious than a simple reduction in discriminative performance: detector rankings depend on the attack family. Because Crescendo was never optimized against \gradsafe{}, this pattern is not evidence of adaptivity; rather, it indicates that an automated context-building attack may avoid producing the local unsafe cues that \mhj{} attacks tend to contain.

The same protocol mismatch shows up for Llama Guard on Crescendo. Table~\ref{tab:crescendo-lg} compares the two prompt-only detectors on the successful subset, where Llama Guard reaches only 0.1478 AUC at three turns and 0.1943 when accumulated. This again says nothing about how Llama Guard would fare if it could read the harmful response; it says that, before generation, both taxonomy classification and gradient scoring find little to hold onto in the user turns of this attack family.

\begin{table}[t]
\centering
\small
\caption{Llama Guard 3 on Crescendo under the same prompt-only, user-turn-only protocol.}
\label{tab:crescendo-lg}
\resizebox{\columnwidth}{!}{
\begin{tabular}{lrrrr}
\toprule
Subset & Method & AUC & AP & TPR@10\%FPR \\
\midrule
Crescendo-all & $W=1$ & 0.2314 & 0.0599 & 0.0200 \\
Crescendo-all & $W=3$ & 0.2614 & 0.0616 & 0.0400 \\
Crescendo-all & Accum. & 0.3435 & 0.0738 & 0.0950 \\
Crescendo-success & $W=1$ & 0.1389 & 0.0425 & 0.0000 \\
Crescendo-success & $W=3$ & 0.1478 & 0.0436 & 0.0000 \\
Crescendo-success & Accum. & 0.1943 & 0.0436 & 0.0000 \\
\bottomrule
\end{tabular}}
	\vspace{-.10in}
\end{table}

\myparatight{Why successful crescendo attacks evade gradient scoring}
\label{sec:crescendo-qual}
The below-random scores, for example, 0.3050 at three turns, indicate more than poor separability: the user turns in successful Crescendo attacks receive lower scores than ordinary benign conversations. To understand this behavior, we examined successful Crescendo conversations. Attackers rarely state the final harmful request directly. Instead, they begin with an informational, historical, fictional, or policy-analysis question and gradually narrow the context until the model supplies unsafe content in its response. We do not reproduce full examples because the relevant evidence lies in the prompt-side structure rather than in operational detail.

This analysis suggests four reasons why a successful Crescendo turn can evade a signature calibrated on single-turn jailbreaks.
\begin{list}{\labelitemi}{\leftmargin=1.2em \itemindent=-0.0em \itemsep=.2em}
    \item \textbf{Neutral, formal phrasing.} The turns use formal and structured language, without the command-like wording, high-perplexity suffixes, or explicit jailbreak markers that the reference signature is designed to capture.

    \item \textbf{No access to assistant responses.} The unsafe content typically appears in the model response, but prompt-only scoring never observes assistant turns.

    \item \textbf{Contextual dilution.} Early Crescendo turns are benign in isolation, so combining them with the turn carrying the harmful intent dilutes the small unsafe cue that remains in the prompt.

    \item \textbf{Overlap with realistic benign traffic.} Genuine WildChat conversations can contain emotionally intense, adversarial-looking, or policy-sensitive discussion, which may induce a stronger safety gradient than the polished Crescendo prompts.
\end{list}

\begin{table}[t]
\centering
\small
\caption{Qwen2.5-7B-Instruct window ablation on \mhj{} versus stratified \wildchat{}. Transfer FPR applies the Llama 3.1 synthetic threshold of 0.0205. Settings cluster near random and accumulated scoring is best, reversing the Llama 3.1 ordering.}
\label{tab:qwen-cross-model}
\resizebox{\columnwidth}{!}{
\begin{tabular}{lrlrrr}
\toprule
Method & AUC & 95\% CI & AP & TPR@10\%FPR & Transfer FPR \\
\midrule
$W=1$ & 0.5613 & [0.5348, 0.5893] & 0.2461 & 0.1453 & 0.8290 \\
$W=2$ & 0.5636 & [0.5368, 0.5891] & 0.2511 & 0.1434 & 0.7510 \\
$W=3$ & 0.5730 & [0.5467, 0.5976] & 0.2478 & 0.1285 & 0.6930 \\
$W=4$ & 0.5653 & [0.5393, 0.5916] & 0.2437 & 0.1248 & 0.6640 \\
$W=5$ & 0.5627 & [0.5366, 0.5883] & 0.2394 & 0.1210 & 0.6560 \\
$W=7$ & 0.5545 & [0.5279, 0.5809] & 0.2336 & 0.1006 & 0.6430 \\
Accum. & 0.5908 & [0.5669, 0.6162] & 0.2458 & 0.0968 & 0.3880 \\
\bottomrule
\end{tabular}}
	\vspace{-.20in}
\end{table}

\myparatight{Cross-model sensitivity}
\label{sec:cross-model}
To see whether these conclusions are tied to one model, we repeat the evaluation on Qwen2.5-7B-Instruct with a separately built reference signature, scoring the same 537 \mhj{} attacks and 2{,}000 \wildchat{} conversations, with no rows dropped for errors. Table~\ref{tab:qwen-cross-model} reports the full sweep, and two things stand out.

First, separability nearly disappears. The best Qwen setting is accumulated scoring at an AUC of 0.5908, while every windowed setting lies between 0.5545 and 0.5730, only slightly above random; by contrast, every Llama 3.1 8B setting exceeds 0.73. Second, the ordering reverses. On Llama, AUC declines monotonically from 0.8101 at one turn to 0.7362 at seven turns, with accumulated scoring performing worst. On Qwen, accumulated scoring performs best, and the one-turn window at 0.5613 is among the weakest settings. Thus, the gradient-dilution explanation observed for Llama does not extend to Qwen. Threshold portability also fails: transferring the Llama 3.1 synthetic threshold of 0.0205 to Qwen produces false-positive rates ranging from 38.8\% for accumulated scoring to 82.9\% for one-turn scoring. This result confirms that thresholds do not transfer reliably across either architectures or benign distributions.

The pattern inside the Qwen sweep reinforces this reversal. The windowed settings stay nearly flat between 0.5545 and 0.5730 rather than declining monotonically as they do on Llama 3.1, and the transfer FPR falls as the window grows, which is again the reverse of Llama 3.1, where every windowed setting produced a similarly high rate.

Together, these results show that the behavior of a gradient detector depends on the model on which it is calibrated rather than being a fixed property of the method. The near-random performance on Qwen is consistent with a differently structured safety-relevant gradient subspace under the same signature procedure. We do not claim that Qwen is fundamentally harder or that the Llama results establish a robust method, because each result reflects one signature realization against one benign benchmark. The key lesson is that thresholds do not transfer across models and must be recalibrated for each target.

\myparatight{Analysis of simple extensions}
Before concluding that the performance drop on realistic benign data is difficult to address, we evaluated three intuitive remedies. Table~\ref{tab:negative} shows that none outperforms the standard three-turn \gradsafe{} baseline. We include these results because each tests a different hypothesis about the source of the failure, and the results do not support these hypotheses.

\begin{table}[t]
\centering
\small
\caption{Intuitive extensions. None beats the plain $W=3$ baseline on the real \wildchat{} benchmark or its pilot subsets.}
\label{tab:negative}
\resizebox{\columnwidth}{!}{
\begin{tabular}{lrrrr}
\toprule
Pilot & Attack $n$ & Benign $n$ & AUC & AP \\
\midrule
Baseline $W=3$ & 537 & 2000 & 0.7614 & 0.4278 \\
Contrastive raw delta & 537 & 2000 & 0.6672 & 0.3094 \\
Contrastive projection delta & 537 & 2000 & 0.7174 & 0.3481 \\
Contrastive ratio & 537 & 2000 & 0.6754 & 0.3149 \\
Subspace pilot baseline & 120 & 120 & 0.7793 & n/a \\
Subspace residual cosine & 120 & 120 & 0.7042 & n/a \\
Multi-turn ref baseline & 20 & 20 & 0.7812 & n/a \\
Multi-turn reference & 20 & 20 & 0.7512 & n/a \\
\bottomrule
\end{tabular}}
\end{table}

The pattern is consistent across all three remedies. In the contrastive variants, subtracting or normalizing a contextual baseline performs worse than the original baseline, indicating that the subtraction removes useful unsafe signal together with benign context. In the subspace variant, estimating and removing benign directions performs worse than its pilot baseline, indicating that benign conversational structure overlaps with the attack signal it was intended to preserve. Building a multi-turn reference direction changes the reference meaningfully but still does not outperform the baseline, showing that re-anchoring alone cannot resolve the overlap introduced by realistic benign traffic. In short, the failure is not attributable to a simple post-hoc adjustment.

\section{Discussion}

\myparatight{The signal remains detectable, but its practical utility is limited}
On realistic traffic, one-turn and two-turn scoring outperform random ranking by a wide margin, and tactic-level AUCs are strong for injection, output-format, and request-framing attacks. However, the results do not indicate that the problem is solved: even the best setting remains far from a deployable operating point once realistic benign conversations are included.

\myparatight{Limited generalization across attack families}%
Our results on Crescendo show that the short-window signal observed on MHJ does not transfer reliably across multi-turn jailbreak families. On MHJ, short-window gradients remain useful despite the realistic-benign-data gap; on successful Crescendo attacks, however, the same scores approach or fall below random performance, and longer windows rank attacks below benign traffic. The most likely explanation is that a context-building attack keeps its user turns close to benign gradients until the model, rather than the user, generates the unsafe content.

\myparatight{More context can hurt}
The ablation is best explained by dilution: a longer window contains more benign or unrelated content, and the gradients induced by that content move the aggregate away from the unsafe reference. Therefore, accumulated full-history scoring is not a safe default even for a long-context model. The fact that a single-turn baseline outperforms every context-aware setting challenges the assumption that more history is automatically safer. This result does not make context useless; rather, it shows that naive concatenation can average a sparse attack signal with benign noise. It motivates context-conditional gating or asymmetric scoring that isolates safety-critical parameters without incorporating this noise.

\myparatight{Realistic benign calibration is not optional}
Synthetic benign data produced both very high scores and overly optimistic thresholds. Real conversations include ambiguous requests, technical instructions, role-play, policy-sensitive material, and formatting that may resemble attacks. Therefore, multi-turn detection must be evaluated against realistic benign distributions, ideally length-matched to the attacks; otherwise, reported performance is unlikely to remain reliable in deployment.

\myparatight{External classifiers answer a different question}
Llama Guard is designed to classify prompts and responses under a taxonomy, and our protocol instead asks it to spot hidden intent before any harmful output exists, using only user turns. Its low AUC here, including on successful Crescendo, is therefore best read not as a defect of the classifier but as confirmation that pre-generation hidden-intent detection is a genuinely different intervention point from response-side moderation.

\myparatight{What this means for practice}
Five rules follow for practitioners. First, on Llama 3.1 8B Instruct the default should be single-turn max-over-turn scoring, which beats three-turn, seven-turn, and accumulated settings, but this choice is model-specific, since accumulated scoring wins on Qwen2.5-7B-Instruct, so the window and the aggregation must be validated per model family. Second, thresholds should be tuned on the deployment's own benign distribution rather than on synthetic prompts, because the synthetic three-turn threshold transferred to 90.85\% FPR on \wildchat{}. Third, long conversations need length-aware thresholds or audits, since more history means more dilution and accumulated scoring is not automatically safer. Fourth, deployments should watch for attack-family drift, because a detector that cleanly separates \mhj{} can still collapse on Crescendo. Fifth, prompt-only detectors should be evaluated apart from response-side moderation, since pre-generation blocking is valuable precisely because it acts before output exists, but for the same reason it cannot use evidence that only the response contains.

\section{Conclusion}

We evaluate gradient-based jailbreak detection in multi-turn dialogue by adapting \gradsafe{} with fixed-size context windows. Although the detector performs well under synthetic benign data, its performance decreases substantially on realistic \wildchat{} conversations, and thresholds calibrated on synthetic data do not transfer. Short windows provide better separability than longer or full-history contexts, but successful Crescendo attacks remain difficult to distinguish from benign conversations. Results also vary across attack-generation methods and target models. These findings show that multi-turn gradient-based detection should be evaluated with realistic benign data, multiple context lengths, diverse attack families, and different model architectures.

\begin{acks}
We thank the anonymous reviewers for their comments.
\end{acks}

\balance

\bibliographystyle{ACM-Reference-Format}
\bibliography{references}

\end{document}